\documentclass[twocolumn,english]{iopart}
\usepackage[T1]{fontenc}
\usepackage[latin9]{inputenc}
\usepackage{amsbsy}
\usepackage{amssymb}

\makeatletter
\usepackage{iopams}
\usepackage{setstack}

\usepackage{bm}% bold math
\usepackage{hyperref}

\makeatother

\usepackage{babel}
\begin{document}

\title{Entanglement susceptibility: Area laws and beyond}

\author{Paolo Zanardi and Lorenzo Campos Venuti}

\address{Department of Physics and Astronomy, and Center for Quantum Information
Science \& Technology, University of Southern California, Los Angeles,
CA 90089-0484}
\begin{abstract}
Generic quantum states in the Hilbert space of a many body system
are nearly maximally entangled whereas low energy physical states
are not; the so-called area laws for quantum entanglement are widespread.
In this paper we introduce the novel concept of entanglement susceptibility
by expanding the 2-Renyi entropy in the boundary couplings. We show
how this concept leads to the emergence of area laws for bi-partite
quantum entanglement in systems ruled by local gapped Hamiltonians.
Entanglement susceptibility also captures quantitatively which violations
one should expect when the system becomes gapless. We also discuss
an exact series expansion of the 2-Renyi entanglement entropy in terms
of connected correlation functions of a boundary term. This is obtained
by identifying Renyi entropy with ground state fidelity in a doubled
and twisted theory.

% and may pave the way to the
%non perturbative regime.
%We find an exact series expansion of the Renyi entanglement
%entropy in terms of connected correlation functions of a boundary
%term. This is obtained by identifying Renyi entropy with ground state
%fidelity in a doubled and twisted theory and may pave the way to the
%non perturbative regime.

%we 
%present an elementary yet general argument for the emergence of area laws
%for bi-partite quantum entanglement in systems ruled by local gapped
%Hamiltonians. This goal is achieved by introducing the perturbative
%notion of entanglement susceptibility and establishing its relation
%to fidelity susceptibility. Entanglement susceptibility also captures
%quantitatively which violations one should expect when the system
%becomes gapless. 

%and how to generalize the analysys of area laws to generic i.e., excited, Hamiltonian eigenstates.

\end{abstract}
\maketitle

\section{Introduction}

The quantum-mechanical state space of a many-body system is huge,
but an exponentially large fraction of it is inhabited by fairly unphysical
states \cite{illusion}. Roughly speaking one may say that, since
we live in a relatively cold universe ruled by local quantum field
theory, the vast majority of states one encounters in physically relevant
situations are highly {\em non} generic as they are low-energy
states of local Hamiltonians.

For example it has been known for some time now that {\em generic}
random quantum states are nearly maximally entangled \cite{generic}
and that this generic property implies a so-called {\em volume law}
for the entanglement entropy $S.$ More specifically partitioning
a system into two regions $A$ and $B$ with $N_{A}$ and $N_{B}$
spins respectively and a with a common boundary $\partial A$, the
entanglement between $A$ and $B$ is $O(\min\{|A|,|B|\}).$ At variance
with this generic prediction, area laws i.e., $S=O(|\partial A|)$,
for quantum entanglement (as well as other information theoretic quantities
\cite{frank}) are ubiquitous in physics: from black hole thermodynamics
to quantum entanglement in zero-temperature many body systems (see
\cite{AL} for a review).

Moreover also violations of these laws have a fundamental physical
meaning, as they signal the divergence of an underlying correlation
length e.g., quantum criticality \cite{luigi} or give rise to sub-leading
terms that allow one to detect hidden topological order \cite{TEE}.
Ensembles of physical random states have recently been shown to feature
an area law in the {\em typical} case \cite{physical}. Finally,
the very fact that area laws hold for many systems of interest i.e.,
the relevant states are just slightly entangled, is the key feature
at the grounds of efficient classical simulations algorithms for quantum
systems \cite{Vidal}.

While locality and finite range correlations for gapped system certainly
provide an intuitive ground for understanding the origin of area laws,
proving them in a rigorous fashion is often quite hard \cite{matt1}.
For the sake of concreteness let us focus on the ground state of a
local Hamiltonian: what makes it different from a generic quantum
states? More specifically, why is an area law for entanglement obeyed
as opposed to a volume law?

The goal of this paper is to show how one can produce a surprisingly
simple argument for the emergence of bi-partite entanglement area
laws for low-energy states e.g., ground states of gapped systems with
local Hamiltonians. The argument we are going to discuss is both quantitative
and general and is based on elementary perturbative expansion of the
2-Renyi entropy with respect the boundary Hamiltonian terms. This
allows one to introduce a natural notion of entanglement susceptibility
that in turn, and in spite of its simplicity, is able to unveil scaling
behavior. Remarkably entanglement susceptibility also shows in which
ways area laws might be violated for gapless systems and excited energy
eigenstates.

In this paper we also discusses an intimate connection between bi-partite
entanglement and ground state fidelity \cite{Fid} as well as a relation
between entanglement susceptibility and fidelity susceptibility \cite{gu,Lor2007}.
%This relation between two quantities is physically
%compelling as both are known to feature singular behavior
%when a systems becomes gapless.
% and therefore can be  used to detect and 
%study quantum phase transitions \cite{sachdev}.
Fidelity and entanglement are two of the most fundamental quantum
information tools and both can be exploited to study quantum phase
transitions \cite{sachdev}. Entanglement shows that at criticality
unique quantum correlations become long-ranged and fidelity witnesses
the ensuing orthogonality catastrophe in the Hilbert space. We find
it conceptually rewarding that these two phenomena appear now as two
sides of the same coin.

\section{Entanglement susceptibility}

Let us consider a spin system with state-space ${\cal H}_{\Lambda}:=\otimes_{i\in\Lambda}h_{i}\cong({\mathbf{C}}^{d})^{\otimes\,|\Lambda|}$
and let $(A,B)$ be a bi-partition of the set of vertices $\Lambda$
such that, with obvious notation, ${\cal H}_{\Lambda}={\cal H}_{A}\otimes{\cal H}_{B}.$
Any local Hamiltonian $H=\sum_{X\subset\Lambda}H_{X}$ over ${\cal H}_{\Lambda}$
($H_{X}\in B({\cal H}_{X}$) can be written as $H=H_{A}+H_{B}+H_{\partial}$
where $H_{A/B}=\sum_{X\subset A/B}H_{X}$ and $H_{\partial}=\sum_{X\cap A\cap B\neq\emptyset}H_{X}.$
We are now interested in the eigen-state properties of the following
family of Hamiltonians associated to $H$: 
\begin{equation}
H(\lambda)=H_{A}+H_{B}+\lambda H_{\partial},\quad(\lambda\in[0,1])\label{Ham}
\end{equation}
 Of course $H(0):=H_{0}$ describes the situation in which the two
regions $A$ and $B$ are decoupled whereas $H(1)$ is nothing but
the original Hamiltonian where boundary coupling $H_{\partial}$ is
assumed to be non-trivial. The eigenstates of $H_{0}$ have the factorized
form $|p_{A}\rangle|p_{B}\rangle$ with eigenvalues $E_{p_{A}}+E_{p_{B}},$
the non-degenerate ground state being $|\Psi_{0}\rangle=|0_{A}\rangle|0_{B}\rangle$
with energy $E_{0}.$ Let $|\Psi_{0}(\lambda)\rangle$ denote the
ground state of (\ref{Ham}) and $\rho_{A}:={\rm {Tr}}_{B}|\Psi_{0}(\lambda)\rangle\langle\Psi_{0}(\lambda)|$
the reduced density matrix of the region $A$. In order to quantify
the entanglement between $A$ and $B$ we use the Renyi 2-entropy
$S_{2}=-\log{\rm {Tr}}\rho_{A}^{2}.$ The first result we would like
to present is: 
\begin{eqnarray}
S_{2}(\lambda) & = & 2\lambda^{2}\chi_{E}+O(\lambda^{3})\nonumber \\
\chi_{E} & := & \sum_{p_{A},p_{B}\ge1}\left|\frac{\langle p_{A}p_{B}|H_{\partial}|0_{A}0_{B}\rangle}{E_{p_{A}}+E_{p_{B}}-E_{0}}\right|^{2}\label{S2}
\end{eqnarray}
 In the remainder of the paper we will refer to this quantity as {\em
entanglement susceptibility} \cite{alpha}. The derivation of (\ref{S2})
is straightforward based on Rayleigh-Schrödinger perturbation theory
and is given in the Appendix. Now a couple of simple qualitative comments:

1) The entanglement susceptibility (\ref{S2}) depends explicitly
on the Hamiltonian term $H_{\partial}$. This dependence makes manifest
the intuitive fact that entanglement between $A$ and $B$ is caused
by interactions terms supported on the boundary between them.

2) The structure of (\ref{S2}) clearly indicates that large gaps
(density of states) above the ground state energy may result in small
(large) value for $\chi_{E}.$ It should also be clear, that vanishing
gaps may induce singular behavior of $\chi_{E}$ depending on the
behavior of the corresponding matrix elements of $H_{\partial}.$
This last fact establishes a direct and transparent connection between
the behavior of entanglement susceptibility and quantum criticality.

\section{The area law bound}

To obtain the area law from (\ref{S2}) we introduce the so-called
fidelity susceptibility \cite{gu,DGQPT} $\chi_{F}:=-\partial^{2}\log{\cal F}(\lambda)\partial\lambda^{2}|_{\lambda=0},$
where ${\cal F}(\lambda):=|\langle\Psi_{0}(\lambda)|\Psi_{0}\rangle|$
is the ground state fidelity. ${\cal F}$ is the main ingredient of
the so called fidelity approach to quantum critical phenomena \cite{Fid}
and has been thoroughly investigated over the last few years. The
basic fact is that: 
\begin{equation}
\chi_{E}\le\chi_{F}\label{boundEF}
\end{equation}
 Indeed $\chi_{F}$ , for the Hamiltonian family (\ref{Ham}) at $\lambda=0,$
is given by $\sum_{p_{A}+p_{B}\ge1}|\frac{\langle p_{A}p_{B}|H_{\partial}|0_{A}0_{B}\rangle}{E_{p_{A}}+E_{p_{B}}-E_{0}}|^{2}$
\cite{gu} that has the very same structure of the term we are willing
to bound but without the restriction that {\em both} $p_{A}$ and
$p_{B}$ are larger than zero \cite{reminder}. The bound (\ref{boundEF})
is not always tight. In fact, in order to have potential singularities
in $\chi_{E}$ the gaps of subsystem $A$ {\em and} $B$ have to
vanish. %\textbf{This is in fact confirmed on the hand of an explicit
%example, see section }\textbf{\emph{Area law corrections}}\textbf{
%below.}
 On the contrary for $\chi_{F}$ the closure of the gap in any of
the two subsystem e.g., say\textbf{ }the largest one is approaching
criticality, may be enough to generate a divergent behavior. %Our main conclusion could be now directly drawn from (\ref{S2}) and
%the results of \cite{Lor2007}, however, for the sake of completeness
%we explicty 
Now we show how the bounds proven in \cite{Lor2007} can be adapted
to the current case to provide an area law. One can write $\chi_{F}=\|G_{0}H_{\partial}|\Psi_{0}\rangle\|^{2}$
where $G_{0}=\sum_{p>0}|p\rangle\langle p|/(E_{p}-E_{0})$, and observe
that for gapped systems $\Delta:=E_{1}-E_{0}>0$ and $\|G_{0}\|:=\sup_{\|\psi\|=1}\|G_{0}\psi\|\le\Delta^{-1}$
from which one has 
\begin{eqnarray}
\chi_{F} & \le & \|G_{0}\|^{2}\|QH_{\partial}|\Psi_{0}\rangle\|^{2}\le\Delta^{-2}\langle\Psi_{0}|H_{\partial}QH_{\partial}|\Psi_{0}\rangle\nonumber \\
 & = & \Delta^{-2}\left(\langle\Psi_{0}|H_{\partial}^{2}|\Psi_{0}\rangle-\langle\Psi_{0}|H_{\partial}|\Psi_{0}\rangle^{2}\right).
\end{eqnarray}
 This last factor is nothing but the {\em connected} correlation
function for the boundary operator $H_{\partial}$ in the ground state
$\Psi_{0}.$ Thanks to the exponential decay of correlations for gapped
ground states of local Hamiltonians \cite{Matt} the connected correlation
above scales as the support of the involved operators. More precisely
if $H_{\partial}=\sum_{j\in\partial A}h_{j}$ then $\left|\langle H_{\partial}H_{\partial}\rangle_{c}\right|\le\xi^{d-1}\max_{j\in\partial A}\|h_{j}\|^{2}|\partial A|,$
where $\xi$ is of the order of the finite correlation length of the
system measured in units of some microscopic length scale e.g., lattice
spacing. Therefore, using (\ref{boundEF}), one gets 
\begin{equation}
\chi_{E}\le\frac{1}{\Delta^{2}}\max_{j\in\partial A}\|h_{j}\|^{2}\xi^{d-1}|\partial A|,\label{AreaBound}
\end{equation}
 %immediately sees that $\chi$ fulfills
%an area law an this in turn implies a similar one for the entanglement susceptibility $\chi_E$.
In words: the entanglement susceptibility for a gapped (non degenerate)
ground state of a local Hamiltonian obeys an area law. %This  area law for the entanglement susceptibility is the main result of this paper.
We would like to emphasize that this argument shows that violations
of the area law i.e., of the bound (\ref{AreaBound}), are a sufficient
condition for criticality (for local Hamiltonians). On the other in
\cite{Lor2007} we showed that in the case in which $H_{\partial}$
has a large scaling dimension criticality does not necessarily imply
super-extensive scaling of fidelity susceptibility. In the present
case this suggests that one may have gapless systems that still obey
an area law. Such phenomena have already been pointed out in the context
of projected entangled pair states in \cite{Frank2006} or for topological
insulators which have gapless edge modes but a finite correlation
length in the bulk \cite{topo-insu}. 

Another interesting point is the validity of these results for {\em
excited} Hamiltonian eigenstates. The derivation of (\ref{S2}) shows
that as long as the excited state is energetically well separated
from the rest of the spectrum (\ref{S2}) holds true. Moreover if
the eigenstate is also clustering i.e., connected correlations functions
of local operators decays sufficiently fast, also the derivation of
(\ref{AreaBound}) is valid and hence an area law fulfilled. The way
this may fail is a high (unperturbed) density of states that makes
the excited state (quasi-)degenerate and/or a lack of clustering.
These remarks are consistent with the findings of \cite{masanes}.

%Now that we have presented our main result (\ref{AreaBound}), and before going back to general considerations, we are
%in the position of illustrating it with a few concrete examples

\section{Example: Quasi-free systems}

The area law prediction of Eq (\ref{AreaBound}) for gapped local
Hamiltonian can be confirmed by explicit, and independent, calculations
for exactly solvable models. In this section, for the sake of illustration,
we discuss two cases:

{\em{a)}} Free fermions on a graph $\Lambda$ with quadratic
Hamiltonian $H=\sum_{i,j\in{\Lambda}}Z_{ij}c_{i}^{\dagger}c_{j}.$
The $c_{i}$'s are canonical fermionic operators i.e. $\{c_{i},c_{j}^{\dagger}\}=\delta_{ij},\,\{c_{i},c_{j}\}=0$
and $Z=(Z_{i,j})_{i,j}$ is a real symmetric $|\Lambda|\times|\Lambda|$
matrix, local on $\Lambda$ \cite{local}. The analog of equation
(\ref{Ham}) in this case is given by $Z(\lambda)=Z+\lambda\delta Z$
where $\delta Z$ is with non vanishing entries on a $O(|\partial A|\times|\partial A|)$
sub-block (rank($\delta Z)=O(|\partial A|).$ The gapfulness assumption
in this single particle picture amounts to the {\em invertibility}
of $Z$ i.e., the minimum $Z$ singular value $\Delta$ is bounded
away from zero. The fidelity susceptibility is known to have the form
$\chi_{F}=\|\partial T/\partial\lambda\|_{2}^{2}$ where $T(\lambda):=Z|Z|^{-1}$
is the unitary part of the polar decomposition of $Z$ and $\|X\|_{2}:=\sqrt{{\rm {Tr}}X^{\dagger}X}$
\cite{sly-prl}. We can now use the bound (see chapter VII.5 of \cite{bhatia})
\[
\left\Vert T_{A}-T_{B}\right\Vert _{2}\le\frac{2}{\left\Vert A^{-1}\right\Vert ^{-1}+\left\Vert B^{-1}\right\Vert ^{-1}}\left\Vert A-B\right\Vert _{2}
\]
(with $\left\Vert X\right\Vert $ indicating the operator norm of
$X$), for $A,B$ invertible operators with unitary factors $T_{A,B}$
in their polar decomposition. Setting $A=Z$ and $B=Z+\delta Z$ in
the above inequality we get $\left\Vert \delta T\right\Vert _{2}\le2/(\Delta+\Delta')\left\Vert \delta Z\right\Vert _{2}$
with $\Delta'$ indicating the smallest singular value of $Z+\delta Z$.
 Moreover $\|\delta Z\|_{2}^{2}\le{\rm {rank}}(\delta Z)\|\delta Z\|^{2}$
with $\|\delta Z\|=O(1).$ Thanks to (\ref{boundEF}) this allows
one to conclude 
\begin{equation}
\chi_{E}\le\frac{4}{(\Delta+\Delta')^{2}}{\rm {rank}}(\delta Z)\|\delta Z\|^{2}=\frac{1}{\Delta^{2}}O(|\partial A|).\label{quasifree}
\end{equation}
 {\em{b)}} Quasi-free bosons. In this case the Hamiltonian is
given by $H=1/2\sum_{i\in\Lambda}p_{i}^{2}+1/2\sum_{i,j\in\Lambda}V_{ij}x_{i}x_{j},$
$[x_{i},p_{j}]=i\hbar\delta_{ij}.$The harmonic coupling matrix $V$
is symmetric, non negative definite, local on $\Lambda$ and $\|V\|=O(1).$
The ground state is known to be a Gaussian state $|V\rangle$ with
covariance matrix $\Gamma(V)=V^{-1/2}\oplus V^{1/2}$ and the gap
with the first excited state is given by $\Delta=2\lambda_{min}(V^{1/2})$
($\lambda_{min}(X):=$ minimum eigenvalue of $X$) \cite{cramer}.
The ground state fidelity can be obtained by performing a simple Gaussian
integral and yields: ${\cal F}=|\langle V|V^{\prime}\rangle|\sim[\det(\Gamma^{-1}+(\Gamma^{\prime})^{-1})]^{-1/4};$
setting $\Gamma^{\prime}:=\Gamma(V+\delta V)=\Gamma+\delta\Gamma$
and expanding at leading order in $\delta\Gamma$ one finds for the
fidelity susceptibility: $\chi_{F}\sim\|\Gamma^{-1}\delta\Gamma\|_{2}^{2}.$
This quantity can be bounded above as follows: $\chi_{F}\le1/\lambda_{min}^{2}(\Gamma)\|\delta\Gamma\|_{2}^{2}=O(\|\delta\Gamma\|_{2}^{2}/\Delta^{2}).$
Moreover, from the expression of $\Gamma(V)$ above, one obtains $\delta\Gamma=1/2(V^{-1/2}\delta V\oplus V^{-3/2}\delta V)$
whence $\|\delta\Gamma\|_{2}=1/2(\|V^{-1/2}\delta V\|_{2}+\|V^{-3/2}\delta V\|_{2})=O(\Delta^{-3/2}\|\delta V\|_{2}).$
Finally, $\chi_{E}\le\chi_{F}=O(\Delta^{-5}{\rm {rank}}(\delta V)\|\delta V\|^{2})=O(|\partial A|).$
In the last steps we exploited Eq.~(\ref{boundEF}) and the fact
that the boundary perturbation $\delta V$ has rank $O(|\partial A|)$
and $O(1)$ norm.

\section{Area law corrections in gapless systems}

In this section we would like to illustrate how the entanglement susceptibility
(\ref{S2}) is able to capture the modifications to the strict area
law that one expects in critical i.e., gapless systems. To this aim
let us consider a $d$-dimensional tight-binding model on a hypercubic
lattice. We partition the system into two half-spaces $A$ and $B$
of width $L_{A}-1$ and $L_{B}-1$ separated by a $(d-1)$-dimensional
hyperplane and use the parametrization $x=(\boldsymbol{x}_{\parallel},x_{\perp})$.
To be consistent we fix the boundary conditions to be periodic along
$\boldsymbol{x}_{\parallel}$ and open along $x_{\perp}$. For simplicity
we fix the size of all the parallel directions to $L$. The boundary
term has the form: $H_{\partial}=V+V^{\dagger}$ where $V:=\sum_{\boldsymbol{x}_{\parallel}\in\partial}c_{\boldsymbol{x}_{\parallel},x_{\perp}^{0}}^{A\dagger}c_{\boldsymbol{x}_{\parallel},x_{\perp}^{0}}^{B}$
and we can set $x_{\perp}^{0}=1$ without loss of generality. The
boundary operator $V$ can then be expressed in terms of the Fourier
basis which diagonalize $H_{A/B}$ and turns out to be $V=\sqrt{4/(L_{A}L_{B})}\sum_{\boldsymbol{k}_{\parallel},k_{\perp},q_{\perp}}\sin(k_{\perp})\sin(q_{\perp})c_{\boldsymbol{k}_{\parallel},q_{\perp}}^{A\dagger}c_{\boldsymbol{k}_{\parallel},k_{\perp}}^{B}$.
Given the form of the one particle dispersion $\epsilon_{k}=2\sum_{i=1}^{d}\cos(k_{i})$
the energy denominator in eq.~(\ref{S2}) depends only on the perpendicular
variables. The entanglement susceptibility takes the form 
\begin{equation}
\chi_{E}=\frac{8}{L_{A}L_{B}}\,\sum_{\boldsymbol{k}_{\parallel},k_{\perp},q_{\perp}}\frac{\sin^{2}(k_{\perp})\sin^{2}(q_{\perp})n_{k}(1-n_{q})}{(\epsilon_{k}-\epsilon_{q})^{2}}\,.\label{eq:chiE}
\end{equation}
 where the occupation numbers are given by $n_{k}=\vartheta(-\epsilon_{\boldsymbol{k}_{\parallel,k_{\perp}}})$
($\vartheta$ is the Heaviside step function) and we used the notation
$q=(\boldsymbol{k}_{\parallel},q_{\perp})$. For large $L$ we can
introduce the function $\Xi(k_{\perp},q_{\perp})=\int d\boldsymbol{k}_{\parallel}\, n_{\boldsymbol{k}_{\parallel},k_{\perp}}(1-n_{\boldsymbol{k}_{\parallel},q_{\perp}})$
analogous to $\Xi(q)$ defined in \cite{wolf}. $\Xi$ measures the
volume of $\boldsymbol{k}_{\parallel}$ such that $(\boldsymbol{k}_{\parallel},k_{\perp})$
lies in the Fermi sea while $(\boldsymbol{k}_{\parallel},q_{\perp}$)
does not. For the specific tight-binding model at half filling $\Xi$
can be computed exactly in terms of a difference of volumes of the
standard simplex. In general, for small $\eta:=q-k$, one has $\Xi\left(k,k+\eta\right)\propto\eta\vartheta(\eta-\sigma_{k}-\sigma_{q})$
where $\sigma_{q/k}=\pi/L_{A/B}$ are the infrared cutoffs in momentum
space. For large sizes one obtains $\chi_{E}=L^{d-1}\int_{\sigma_{k}}^{\pi}dk\,\int_{\sigma_{k}+\sigma_{q}}^{\pi-k}d\eta\, f\left(k,k+\eta\right)$.
Expanding $f\left(k,k+\eta\right)=f_{-1}\left(k\right)/\eta+\sum_{n=0}^{\infty}f_{n}\left(k\right)\eta^{n}$
one realizes that the only divergent contribution is given by the
integration of $f_{-1}(k)/\eta$, while all other terms are convergent
as $\sigma_{k/q}\to0$. The integration over $\eta$ gives trivially
a log, and the diverging term turns out to be $\chi_{E}\sim-\alpha_{d}L^{d-1}\ln(\sigma_{k}+\sigma_{q})$
where the constant is given by $\alpha_{d}=\int_{0}^{\pi}f_{-1}\left(k\right)dk$
\cite{function}. If {\em both} $L_{A}$ and $L_{B}$ scale with
$L$ we obtain, for large $L$, 
\begin{equation}
\chi_{E}\sim L^{d-1}\ln L\,.\label{eq:chiE_final}
\end{equation}
 This result is consistent with the logarithmic violations to the
area law for fermionic systems discussed in \cite{wolf} and \cite{klich}.
We would like to stress here the two essential ingredients needed
to obtain Eq.~(\ref{eq:chiE_final}): i) linearity of $\Xi\left(k,k+\eta\right)$
for small $\eta$, more precisely $\Xi\left(k,k+\eta\right)\propto\eta\vartheta\left(\eta-\sigma_{k}-\sigma_{q}\right)$;
and ii) singularity of the form $\left(q-k\right)^{-2}$ for the energy
denominators in Eq.~(\ref{eq:chiE}). These features are believed
to be general for free systems and rely on the existence of the Fermi
surface and the linearity of the particle-hole excitations close to
the Fermi energy. Similar assumptions have been used also in \cite{wolf}.
We thus expect Eq.~(\ref{eq:chiE}) to be valid in more general cases
with more complicated geometry and band structures, at least when
the above assumptions are satisfied.

\section{Beyond the perturbative regime}

Low energy scaling behavior of a physical quantity is a property of
the universality class (in the renormalization group sense) of the
Hamiltonian. As long as the perturbation is not strong enough to induce
a quantum phase transition scaling behavior is not expected to change.
In other words, even though our argument for the entanglement area
law relies on a perturbative expansion we expect it to hold true beyond
the deep perturbative regime i.e., all the way up to $\lambda=1$
in Eq (\ref{Ham}). %Strictly speaking, in order to show that $S_{2}(\lambda=1)$
%obeys an area law, one should prove a) the area law for all higher-order
%entanglement susceptibilities $\chi_{E}^{(n)}:=\partial^{n}S_{2}(0)/\partial\lambda^{n}$
%b) the series $\sum_{n=0}^{\infty}\lambda^{n}\chi_{E}^{(n)}$ as a
%finite convergence radius and can be analytically continued up to
%$\lambda=1.$. 
To elaborate further on this point let us for the moment come back
to consider the ground state fidelity. The quantity $\log{\cal F}$
between the two ground states of the regions $A$ and $B$ when the
boundary coupling $H_{\partial}$ is switched on and off has been
considered in \cite{dubail} and named bi-partite logarithmic fidelity.
The authors of \cite{dubail} provide a heuristic argument, based
on the standard quantum mechanics- classical statistical mechanics
correspondence, to show that the bi-partite logarithmic fidelity fulfills
an area law. The results discussed in this paper so far are tantamount
to a rigorous proof of that claim to the second order in $\lambda.$

To investigate higher order contributions let us notice that for a
gapped system with non-degenerate ground state one has ${\cal F}(\lambda)=\lim_{\beta\to\infty}N(\beta)/N^{1/2}(2\beta),\, N(\beta):=\langle e^{-\beta H(\lambda)}\rangle$
where $\langle\cdots\rangle$ denotes the quantum-mechanical average
over $|\Psi_{0}(0)\rangle$ \cite{non-vanish}. Using the standard
interaction picture formula $e^{-\beta H}=e^{-\beta H_{0}}T_{s}\exp\left(-\lambda\int_{0}^{\beta}dsH_{\partial}(s)\right),\, H_{\partial}(s):=e^{sH_{0}}H_{\partial}e^{-sH_{0}}$
one can prove \cite{kubo} 
\begin{eqnarray*}
 &  & \langle T_{s}\exp\left(-\lambda\int_{0}^{\beta}dsH_{\partial}(s)\right)\rangle=\exp\left(\sum_{n=1}^{\infty}(-\lambda)^{n}c_{n}(\beta)\right)\\
 &  & c_{n}(\beta)=\int_{0}^{\beta}ds_{1}\cdots\int_{0}^{s_{n-1}}ds_{n}\langle H_{\partial}(s_{1})\cdots H_{\partial}(s_{n})\rangle_{c}
\end{eqnarray*}
 where the subscript $c$ denotes {\em connected} (imaginary time)
correlations functions. One can then write: 
\begin{equation}
\log{\cal F}(\lambda)=\lim_{\beta\to\infty}\sum_{n=2}^{\infty}(-\lambda)^{n}\left(c_{n}(\beta)-\frac{1}{2}c_{n}(2\beta)\right).\label{beta}
\end{equation}
 The presence of a spectral gap in $H_{0}$ implies that connected
averages scale as $\beta$ or more precisely $c_{n}(\beta)=A_{n}\beta+B_{n}+O(e^{-\beta\Delta})$
($A_{n},B_{n}$ are $\beta$ independent)\cite{kubo}; whence $\log{\cal F}(\lambda)=\frac{1}{2}\sum_{n=2}^{\infty}\left(-\lambda\right)^{n}B_{n}.$
The term $B_{2}$ is easily seen to yield the fidelity susceptibility
($B_{2}=-\chi_{F}$) and therefore, as we have seen, it is upper bounded
by $|\partial A|$ in the gapped case. %One may conjecture that {\em all} high-order terms are $B_{n}$ are $O(|\partial A|). $ 

The key point is now that all these remarks concerning fidelity are
{\em directly} relevant to Renyi entropy itself. Indeed it turns
out that Renyi 2-entropy {\em is} a particular instance of logarithmic
ground state fidelity \cite{cardy}. To see this fact let us write
the purity in the form ${\rm {Tr}}\rho_{A}^{2}={\rm {Tr}}\left[S_{13}|\Psi_{0}(\lambda)\rangle\langle\Psi_{0}(\lambda)|^{\otimes\,2}\right]$
where $S_{13}$ is the swap operator between the first and third factor
in $({\cal H}_{A}\otimes{\cal H}_{B})^{\otimes\,2}$ i.e., the state
space of two copies of the system \cite{ep}. This trace can be viewed
as the ground state fidelity between the ground state $|\Psi_{0}(\lambda)\rangle^{\otimes\,2}$
of $H^{(2)}:=H(\lambda)\otimes\mathbf{1}+\mathbf{1}\otimes H(\lambda)$
and $S_{13}|\Psi_{0}(\lambda)\rangle^{\otimes\,2}$ i.e., the ground
state of $S_{13}H^{(2)}S_{13}$ \cite{cardy}. %\begin{equation}
%{\cal F}_{twist}=e^{-S_2}
%\end{equation}
Analog constructions extend to all $\alpha$-entropies with integer
$\alpha$ \cite{cardy}. %Now the state-space is doubled and the boundary perturbation
%term is $V_\partial:=H_{\partial}^{(2)}-S_{13}H_{\partial}^{(2)}S_{13}$ .
%Writing ${\cal F}_{twist}=e^{-S_2}$ one s
The fidelity formula (\ref{beta}) shows that one has a direct representation
of the Renyi entropy in terms of connected (imaginary time) correlation
functions of the boundary perturbation $V_{\partial}:=H_{\partial}^{(2)}-S_{13}H_{\partial}^{(2)}S_{13}$
in a {}``doubled and twisted\textquotedbl{} theory. In particular
all the entanglement susceptibilities $\chi_{E}^{(n)}:=\partial^{n}S_{2}(0)/\partial\lambda^{n}$
can be expressed by analogous $B_{n}^{twist}$ terms as defined above,
associated with $V_{\partial}$ \cite{alternative}. Note that, for
finite range interactions, $V_{\partial}$ has $O\left(\left|\partial A\right|\right)$
number of terms and in general represents a surface term. Hence extending
to this doubled theory the heuristic statistical mechanics argument
of \cite{dubail} suggests that the area law should indeed hold for
all the $\alpha$-Renyi entropies $S_{\alpha}(\rho)=-(\alpha-1)^{-1}{\rm Tr}\log\rho^{\alpha},\,\alpha\in\mathbf{N}$)
in the gapped case with clustering \cite{alpha1}. A deeper analysis
of this argument may pave the way towards the rigorous understanding
of how area laws for quantum entanglement and their violations arise
in physics \cite{rigor}.

% would amount to a rigorous and general
%proof of the area law for $S_{2}.$  
%proof of the {}``area law for fidelity\textquotedbl{}.
%%In particular all  entanglement susceptibilities $\chi_{E}^{(n)}:=\partial^{n}S_{2}(0)/\partial\lambda^{n}$
%are given by
%one obtains an alternative and straightforward
%way to derive the entanglement susceptibility formula (\ref{S2})
%as this is just the fidelity susceptibility of the {}``doubled and
%twisted\textquotedbl{} theory. 

\section{Conclusions}

Area laws for quantum entanglement are one of the most common features
showing the strong {\em atypicality} of physical quantum states
in the Hilbert space. In this paper we introduced a natural perturbative
object, the entanglement susceptibility (\ref{S2}). Roughly speaking,
this quantity measures the rate of entanglement generation (as measured
by the Renyi 2-entropy) when an infinitesimal boundary interaction
term is switched on between two previously decoupled regions. Entanglement
susceptibility while much simpler than entanglement entropy itself
is able to describe its scaling behavior. This can be seen by bounding
entanglement susceptibility with fidelity susceptibility \cite{Fid,DGQPT}.
Both area law bounds for gapped systems and possible corrections for
critical ones can be then obtained by elementary means. We illustrated
the area law bounds for entanglement susceptibility with explicit
calculations for the quasi-free systems and logarithmic corrections
in the free fermionic critical case.

The non-perturbative regime can also be explored by realizing that
Renyi entropy is a particular instance of ground state fidelity in
a doubled space \cite{cardy}. This unification allows one to express
entanglement by an exact series expansion of integrated connected
correlations of a twisted boundary interaction.

Let us conclude by mentioning a couple of natural goals for future
research. On the experimental side, very recent work indicates that
entanglement susceptibility may be directly measurable in bosonic
optical lattices \cite{measure}. On the theoretical side, we would
like to extend the results of this paper to excited energy eigenstates
of local Hamiltonians \cite{pasquale}: Are they going to be more
alike typical quantum states?

%and we find it intriguing that they appear to have
%a non trivial interconnection. 

%We believe that a deeper analysis of
%this connection may pave the way towards the understanding of how
%area laws for quantum entanglement and their violations arise in physics.
%A natural goal for future research is to extend the results of this
%paper to general i.e., excited, energy eigenstates.

{\em Acknowledgments} The authors acknowledge partial support by
the ARO MURI grant W911NF-11-1-0268. PZ also acknowledges partial
support by NSF grants No. PHY-969969 and No. PHY-803304. We would
like to thank H. Saleur, S. Haas, A. Hamma and D. Lidar for useful
comments.

\section*{References}{}

\appendix

\section*{Appendix}

{\em{Proof of (\ref{S2}).}} Let us expand the ground state of
(\ref{Ham}) in powers of $\lambda$ : $|\Psi_{0}(\lambda)\rangle={\cal N}\sum_{n=0}^{\infty}\lambda^{n}|\Psi^{(n)}\rangle$
($|\Psi^{(0)}\rangle=|0_{A}\rangle|0_{B}\rangle.$) %where ${\cal N}$ is the normalization.  $|\Psi^{(0)}\rangle=|0_A\rangle|0_B\rangle$ is the {\em factorized} ground state of $H_A+H_B.$  
The perturbative eigenvector corrections $|\Psi^{(n)}\rangle$ are
given by the elementary perturbation theory: $|\Psi^{(n)}\rangle=G_{0}(E_{0})(H_{\partial}|\Psi^{(n-1)}\rangle-\sum_{k=0}^{n-1}E^{(n-k)}|\Psi^{(k)}\rangle);$
where $E^{(k)}$=k-th order correction to the unperturbed eigenenergy
$E^{(0)}=E_{0},$ $G_{0}:=\sum_{n>0}(E_{0}-E_{n})^{-1}|\Psi_{n}\rangle\langle\Psi_{n}|=Q(E_{0}-H_{0})^{-1}Q$
is the (projected) resolvent of $H_{0}$ ($Q:=1-|\Psi_{0}\rangle\langle\Psi_{0}|$.)
%Notice that all the eigenstates of $H_0$ have the form $|p_A,p_B\rangle.$
The reduced density matrix is given by $\rho_{A}(\lambda)={\cal N}^{2}\sum_{n=0}^{\infty}\lambda^{n}\rho^{(n)}$
where $\rho^{(n)}:=\sum_{k=0}^{n}{\rm {Tr}}_{B}|\Psi^{(n-k)}\rangle\langle\Psi^{(k)}|.$
Let us now consider the leading contributions $n=0,1$; % one can write
%$|\Psi_0(\lambda)\rangle ={\cal N}(\lambda)(|\Psi^{(0)}\rangle +\lambda |\Psi^{(1)}\rangle),$ where 
%and  $|\Psi^{(1)}\rangle=\sum_{p_A+p_B>0} C(p_A,p_B) |p_A p_B\rangle,$ where $C(p_A,p_B)=\langle p_A p_B| H_\partial|0_A0_B\rangle/(E_{p_A}+E_{p_B}-E_0).$ The %normalization factor
%at this order is given by ${\cal N}^{-2}=1+\lambda^2 \sum_{p_A+p_B>0} |C(p_A,p_B)|^2.$
performing the partial trace over the region $B$ one finds %$\rho_A(\lambda) =
%{\cal N}^2\, \rm{Tr}_B\left( |\Psi_0\rangle\langle\Psi^0|  +\lambda |\Psi^0\rangle\langle\Psi^1| +  \lambda  |\Psi^{(1)}\rangle\langle\Psi_0| +\lambda^2 |\Psi^1\rangle\langle\Psi^1|\right)=
%\cal N}^2 (\rho_0+\lambda \rho_1+\lambda^2 \rho_2)$ where 
\begin{eqnarray}
\rho^{(0)} & = & |0_{A}\rangle\langle0_{A}|,\quad\rho^{(1)}=\sum_{p_{A}>0}C(p_{A},0)|0_{A}\rangle\langle p_{A}|+{\rm {h.c},}\nonumber \\
\rho^{(2)} & = & \sum_{p_{A},q_{A}}|p_{A}\rangle\langle q_{A}|\sum_{p_{B}}{C(p_{A},p_{B})}\overline{C(q_{A},p_{B})}.
\end{eqnarray}
 %the $|0_A\rangle\langle 0_A|$ term of the last term can be absorbed in to the first
%$\rho_2\to\rho_2^\prime=\rho_2-{\cal N}_0 |0_A\rangle\langle 0_A,|\; \rho_0\to\rho_0^\prime ={\cal N}_0 |0_A\rangle\langle 0_A|$ where ${\cal N}_0:=1+\lambda^2 \sum_{p_B>0} |C(0,p_B)|^2,$
in which $C(p_{A},p_{B})=\langle p_{A}p_{B}|H_{\partial}|0_{A}0_{B}\rangle/(E_{p_{A}}+E_{p_{B}}-E_{0}).$
Let us now compute the purity of $\rho_{A}$: ${\rm {Tr}}\rho_{A}^{2}={\cal N}^{4}{\rm Tr}\left((\rho^{(0)})^{2}+\lambda^{2}(\rho^{(1)})^{2}+\lambda^{2}\{\rho^{(0)},\rho^{(2)}\}\right)+O(\lambda^{3}).$
Here one has to notice that ${\rm Tr}\{\rho^{(0)},\rho^{(1)}\}=0,$
${\rm Tr}\{\rho^{(0)},\rho^{(2)}\}=2\sum_{p_{B}>0}|C(0,p_{B})|^{2},$
and ${\rm Tr}(\rho^{(1)})^{2}=2\sum_{p_{A}>0}|C(p_{A},0)|^{2}.$ The
normalization factor at this order is given by ${\cal N}^{-2}=\||\Psi_{0}\rangle+|\Psi^{(1)}\rangle\|^{2}=1+\lambda^{2}\sum_{p_{A}+p_{B}>0}|C(p_{A},p_{B})|^{2}.$
Using this explicit form and expanding again at leading order one
obtains %\begin{equation}
${\rm {Tr}}\rho_{A}^{2}=1-2\lambda^{2}\sum_{p_{A},p_{B}\ge1}|C(p_{A},p_{B})|^{2}+O(\lambda^{3})$.
Finally taking the negative of log one finds, %for the Renyi 2-entropy $S_2=-\log {\rm{Tr}}\rho_A^2,$
 the expression (\ref{S2}). $\hfill\Box$

This derivation shows that the appearance of the fidelity susceptibility
in our entanglement argument while remarkable is certainly not surprising.
Indeed an important ingredient in the derivation above is the the
normalization factor $1/{\cal N}^{2}=\|\sum_{n=0}^{\infty}\lambda^{n}|\Psi^{(n)}\rangle\|^{2}.$
%\sum_{n=0}^\infty \lambda^n \sum_{k=0}^n \langle\Psi^{(n-k)}|\Psi^{(k)}\rangle.$ 
Since in standard Rayleigh-Schr\"odinger perturbation theory one
can always choose $\langle\Psi^{(0)}|\Psi^{(k)}\rangle=0$ for $k>0,$
it follows that ${\cal N}^{2}$ is nothing but the ground state fidelity:
${\cal F}^{2}=|\langle\Psi_{0}(0)|\Psi_{0}(\lambda)\rangle|^{2}=|\langle\Psi^{(0)}|\Psi_{0}(\lambda)\rangle|^{2}={\cal N}^{2}|\sum_{n=0}^{\infty}\langle\Psi_{0}|\Psi^{(k)}\rangle|^{2}={\cal N}^{2}.$
%The higher order corrections to $|\Psi_{0}(\lambda)\rangle\langle\Psi_{0}(\lambda)|$
%merely give rise to a slight renormalization of $\log{\cal F}$ in
%the calculation of $S_{2}(\lambda)$ i.e., the restriction of the
%sum in (\ref{S2}). Roughly $S_{2}\approx-2\log{\cal F}.$ 


\begin{thebibliography}{References}
\bibitem{illusion} D. Poulin, A. Qarry, R. D. Somma, F. Verstraete,
Phys. Rev. Lett. \textbf{{106}}, 170501(2011)

\bibitem{generic} P. Hayden, D. W. Leung, A. Winter, Comm. Math.
Phys. \textbf{265}, 95-117, (2006)%\bibitem{winter} S. Popescu, A. J. Short, and A. Winter, Nature Physics
%2, 754 (2006); N. Linden, S. Popescu, A. J. Short, and A. Winter,
%Phys. Rev. E \textbf{{79}}, 061103 (2009). 


\bibitem{frank} M.M. Wolf, F. Verstraete, M.B. Hastings, J.I. Cirac,
Phys. Rev. Lett. \textbf{{100}}, 070502 (2008)

\bibitem{AL} J. Eisert, M. Cramer and M. B. Plenio Rev. Mod. Phys.
\textbf{{82}}, 277 (2010)

\bibitem{luigi} L. Amico, R. Fazio, A. Osterloh, V. Vedral, Rev.
Mod. Phys.80, \textbf{{517}} (2008)

\bibitem{TEE} A. Hamma, R. Ionicioiu, and P. Zanardi, Phys. Lett.
A \textbf{{337}}, 22 (2005); A. Kitaev and J. Preskill, Phys. Rev.
Lett. \textbf{{96}}, 110404 (2006); M. Levin and X.-G. Wen, {\em
ibid.} 110405 (2006).

\bibitem{physical} A. Hamma, S. Santra and P. Zanardi, Phys. Rev.
Lett. \textbf{109}, 040502 (2012); A. Hamma, S. Santra and P. Zanardi,
Phys. Rev. A \textbf{86}, 052324 (2012) 

\bibitem{Vidal} G. Vidal, Phys. Rev. Lett. \textbf{{91}}, 147902
(2003)

\bibitem{matt1} For example there exist states with finite correlation
length but exponentially large entanglement: M.~Hastings, Phys. Rev.
B \textbf{{76}}, 035114 (2007)

\bibitem{Fid} P. Zanardi and N. Paunkovi\'{C}, Phys. Rev. E \textbf{{74}},
031123 (2006); H. Q. Zhou and J. P. Barjaktarevic, J. Phys. A: Math.
Theor. \textbf{{41}} 412001 (2008).

\bibitem{gu} W-L. You, Y-W Li, S-J. Gu, Phys. Rev. E \textbf{76},
022101 (2007)

\bibitem{Lor2007} L. Campos Venuti and P. Zanardi, Phys. Rev. Lett.
\textbf{{99}}, 095701 (2007).

\bibitem{sachdev} S. Sachdev, Quantum Phase Transitions (Cambridge
University Press, (1999)

\bibitem{alpha} In the case of a multi parameter boundary coupling
$H_{\partial}:=\sum_{\mu}\lambda_{\mu}V_{\mu}$ the entanglement susceptibility
can be straightforwardly generalized to a tensor $(\chi_{E})_{\mu\nu}=\partial^{2}S_{2}(0)/\partial\lambda_{\mu}\partial\lambda_{\nu}.$
This is a degenerate metric tensor over the parameter manifold. %This is an important result as it allows one to derive almost directly the area law for $S_2.$  


\bibitem{DGQPT} P. Zanardi, P. Giorda and M. Cozzini, Phys. Rev.
Lett. \textbf{{99}}, 100603 (2007)

\bibitem{reminder} With the notation $H_{\partial}^{A/B}\left(\tau\right)=e^{\tau H_{A/B}}H_{\partial}e^{-\tau H_{A/B}}$
and $\langle\cdot\rangle_{c}$ denoting connected correlations on
the ground state, one can also show that $\chi_{E}=\chi_{F}-\int_{0}^{\infty}d\tau\,\tau\left[\langle H_{\partial}^{A}\left(\tau\right)H_{\partial}\rangle_{c}+\langle H_{\partial}^{B}\left(\tau\right)H_{\partial}\rangle_{c}\right]$.

\bibitem{Matt} M. B. Hastings, Phys. Rev. Lett. \textbf{{93}} 140402(2004)

\bibitem{Frank2006} F. Verstraete, M.M. Wolf, D. Perez-Garcia and
J.I. Cirac, Phys. Rev. Lett. \textbf{{96}}, 220601 (2006)

\bibitem{topo-insu} M. Z. Hasan and C. L. Kane, Rev. Mod. Phys.,
\textbf{82}, 3045\textendash{}3067, (2010) 

\bibitem{masanes}L.~Masanes, Phys.~Rev.~A \textbf{80}, 052104
(2009). 

\bibitem{non-vanish} Here we assumed that the overlap of $\langle\Psi_{0}(\lambda)|\Psi_{0}(0)\rangle$
is non vanishing (for finite system size).

\bibitem{dubail} J. Dubail, J-M. Stephan, J. Stat. Mech. L03002 (2011)

\bibitem{ep} P. Zanardi, C. Zalka and L. Faoro. Phys. Rev. A \textbf{{62}},
030301 (2000)

\bibitem{kubo} R. Kubo, J. Phys. Soc. Jpn. \textbf{17}, 1100\textendash{}1120
(1962)

\bibitem{cardy} J. Cardy, Phys.Rev.Lett. \textbf{{106}}, 150404
(2011)

\bibitem{alternative} For example: computing the fidelity susceptibility
of the doubled and twisted theory gives an alternative and straightforward
way of proving (\ref{S2}).

\bibitem{local} By this we mean that the, for all $i\in\Lambda$
the sets ${\cal V}_{i}=\{j\in\Lambda\,/\, V_{ij}\neq0\}$ have size
$O(1).$

\bibitem{bhatia} R.~Bhatia, \emph{Matrix Analysis, }Springer (1997)

\bibitem{sly-prl} S. Garnerone, N. T. Jacobson, S. Haas and P. Zanardi,
Phys. Rev. Lett. \textbf{{102}}, 057205 (2009)

\bibitem{cramer} M. Cramer and J. Eisert, New.~J.~Phys. \textbf{8},
71 (2006)

\bibitem{function} In the present case $f_{-1}\left(k\right)=2/[\pi^{d+1}(d-2)!]\,\left(\pi-k\right)\sin\left(k\right)^{2}$
and $\alpha_{1}=2/\pi^{2}=0.202$, $\alpha_{2}=1/\pi^{2}=0.101$,
$\alpha_{3}=1/(2\pi^{2})=0.05$.

\bibitem{wolf} M.M. Wolf, Phys. Rev. Lett. \textbf{{96}}, 010404
(2006)

\bibitem{klich} D.~Gioev and I.~Klich, Phys.~Rev.~Lett. \textbf{{96}},
100503 (2006) 

\bibitem{alpha1} Notice that also the entanglement susceptibility
notion can be easily extended to the general $\alpha$-Renyi entropy
$S_{\alpha}$ %(\rho)=-(\alpha-1)^{-1}{\rm Tr}\log\rho^{\alpha},$
for $(\alpha\in(1,+\infty)).$ In this case one can easily show that
$S_{\alpha}(\rho_{A}(\lambda))=-\frac{\alpha}{\alpha-1}\lambda^{2}\chi_{E}+O(\lambda^{3}).$
When $\alpha\to1^{+}$ the $\alpha$-Renyi entropy converges to the
von Neumann entanglement entropy $S(\rho):=-{\rm {tr}}\rho\log\rho$
but $d^{2}S_{\alpha}(0)/d\lambda^{2}=-\alpha(\alpha-1)^{-1}\chi_{E}$
diverges to $-\infty.$ The susceptibility is {\em not} defined
for the von Neumann as this latter is not differentiable at pure states.

\bibitem{rigor} One should prove that all the $B_{n}^{twist}$'s
are $O(|\partial A|)$ and that the series $\frac{1}{2}\sum_{n=2}^{\infty}(-\lambda)^{n}B_{n}^{twist}=S_{2}(\lambda)$
has a finite convergence radius.

\bibitem{measure} A. J. Daley, H. Pichler, J. Schachenmayer, P. Zoller,
Phys. Rev. Lett. \textbf{109}, 020505 (2012)

\bibitem{pasquale} V.~Alba, M.~Fagotti and P.~Calabrese, J.~Stat.~Mech.,
P10020 (2009) 

\end{thebibliography}
\end{document}